\documentclass[twocolumn,aps,showpacs]{revtex4}
\usepackage{graphicx}
\newcommand{\be}{\begin{equation}}
\newcommand{\ee}{\end{equation}}
\newcommand{\ben}{\begin{eqnarray}}
\newcommand{\een}{\end{eqnarray}}
\newcommand{\bb}{\bibitem}
\begin{document}

\title{Domain walls in three-field models} 
\author{D. Bazeia$^1$, L. Losano$^1$, and C. Wotzasek$^2$} 
\affiliation{$^1$Departamento de F\'\i sica, Universidade Federal da
Para\'\i ba, Caixa Postal 5008, 58051-970 Jo\~ao Pessoa, PB, Brazil\\
$^2$Instituto de F\'\i sica, Universidade Federal do Rio de Janeiro,
Caixa Postal 68528, 21945-970 Rio de Janeiro, RJ, Brazil}

\date{\today}

\begin{abstract}
We investigate the presence of domain walls in models described by three
real scalar fields. We search for stable defect structures which minimize
the energy of the static field configurations. We work out explict orbits
in field space and find several analytical solutions in all BPS sectors,
some of them presenting internal structure such as the appearence of defects
inside defects. We point out explicit applications in high energy physics,
and in other branches of nonlinear science.
\end{abstract}


\maketitle

This paper deals with domain walls presenting internal structure that
may engender nontrivial phenomena. Specifically we propose new systems
of three coupled real scalar fields whose
dynamics are controlled by superpotentials in the
bosonic sector of supersymmetric theories. The systems allow for
a multiplicity of solitonic solutions including the presence of defects
inside defects, introducing new results that lead to a variety of
applications in diverse fields of physics, chemistry and biology.

Domain walls may involve energy scales as different as those found
in fields of magnetic systems \cite{ms}, nonequilibrium
thermodynamics \cite{wal} and cosmology \cite{co}. 
Usually, domain walls are immersions in $(3,1)$ dimensions of linear
kink-like defects living in $(1,1)$ dimensions. These extended
spatial structures are the basic building blocks to describe the
coexistence of different domains or phases, and they are usually
modeled by a set of scalar fields playing the role of order parameters
satisfying amplitude equations. They also appear in models that
include supersymmetry \cite{susy1,susy2,susy3,susy4}
and supergravity \cite{sugra1,sugra2}. In particular, a great deal of
attention has been given to the restrictions imposed by supersymmetry
in the classification and solution of the solitonic field equations --
in a supersymmetric theory one can categorize the topological defects as
BPS and non-BPS states, according to the work of Bogomol'nyi and of Prasad
and Sommerfield \cite{bps}.

In models described by a single real scalar field we may distinguish at
least two classes of systems, one that supports a single type of defect,
and the other, that supports two or more different defects. We exemplify
this by recalling the $\phi^4$ model, which supports a single type of defect,
the tanh-like kink, and the double sine-Gordon model, which may support two
different defects, the large and the small kinks \cite{01}. On the other hand,
models containing two or more real scalar fields give rise to at least two
other classes of systems - those that support defects that engender internal
structure \cite{is1,is2} and, those that support junctions of
defects \cite{susy1,susy3,susy4}. In the latter context, in Ref.~{\cite{00}},
one has investigated the possibility of a regular hexagonal network
of defects to spring in a model described by two real scalar fields.
In Ref.~{\cite{00a}} the idea of an hexagonal network of defect
to be nested {\it inside} a topological defect has been investigated
in a model described by three real scalar fields. That investigation
has shown that when the host domain wall is driven to relax to cylindrical
or spherical shape, the nested hexagonal network could give rise to
structures resembling nanotubes or fulerenes, respectively. Similar ideas
have been recently presented in \cite{bb}, in the case of a soliton star
that entraps a fulerenelike network of domain walls, and in \cite{bs},
where the Skirme model is used to show the presence of fulerenelike
structures in light nuclei, and also in \cite{ff}, in the case of solutions
of the Einstein equations describing black holes pierced by cosmic strings,
forming structures that very much resemble the Platonic solids.

In Refs.{\cite{00,00a}} one has given up supersymmetry,
to circumvent issues concerning stability of the nested network
of domain walls. Here, however, we return to the basic problem,
which concerns investigating models described by three real scalar fields
in the bosonic portion of supersymmetric theories. Models with three real
scalar fields have been studied before, for instance in an $O(3)$ model
which admits unstable embedded walls \cite{ab}, and also in \cite{agg},
which investigates stability of kinks in a deformed $O(3)$ linear sigma
model. Other motivations include the possibility of building spatial
junctions of defects, instead of the planar junctions considered
in \cite{susy3,susy4}, and the issue of finding a host domain wall
that entraps a network of defects within the framework of supersymmetry.
To construct spatial junctions of defects one first recalls the simpler
issue, of contructing planar junctions. As we know,
we can tile the plane with regular polygons in three distinct ways:
with triangles, squares or hexagons. These cases require
junctions with six, four and three legs, respectively. Thus, the two-field
model should engender discrete $Z_6$, $Z_4$, or $Z_3$ symmetry. Interestingly,
however, we notice that the triangular lattice is dual to the hexagonal
lattice, and vice versa, and that the squared lattice is self-dual. This
means that there is a dual relation between the (triangular, square or
hexagonal) disposition of minima in the two-field model, and the (hexagonal,
square or triangular) lattice made with the junction that spring with the
very same minima. We take this as a guide to construct spatial junctions
of defects in three-field models, to fill space with regular polyhedra.
Before investigating such issue, however, let us first deal with the more
general problem of finding topological defects in three-field models
engendering discrete symmetry.

In general we are interested in potentials that can be obtained in terms
of superpotentials, in the form
\be
V=\frac12W^{2}_{\phi}+\frac12W^{2}_{\chi}
+\frac12W^{2}_{\rho}
\ee
where $W_{\phi_i}=\partial W/\partial\phi_i$, $\phi_i=\phi,\chi,\rho$. 
The energy density for static solutions that depend on the single spatial
coordinate $x$ can be written as
\be
{\cal E}(x)=\frac{dW}{dx}+\sum_i \left(\frac{d\phi_i}{dx}-
W_{\phi_i}\right)^2
\ee
It can be minimized to give $E_{ab}=|\Delta W_{ab}|$, with
\be
\Delta W_{ab}=W_a-W_b
\ee
where the indices identify vacuum states, such that
$W_a=W({\bar\phi}_a,{\bar\chi}_a,{\bar\rho}_a)$ and so forth.
This energy is the energy of Bogomol'nyi-Prasad-Sommerfield (BPS)
solutions \cite{bps,b1}. The BPS states obey the first order ordinary
differential equations 
\be
\frac{d\phi}{dx}=W_{\phi},\;\;\;\;\;
\frac{d\chi}{dx}=W_{\chi},\;\;\;\;\;
\frac{d\rho}{dx}=W_{\rho}
\ee

We now introduce the model, given by the superpotential depending
on two real parameters, $r$ and $s$,
\be
\label{w2}
W=\phi-\frac13\phi^3-r\phi(\chi^2+
\rho^2)+rs\rho^2
\ee
Here we are using dimensionless space-time coordinates and fields.
The corresponding potential is
\ben
\label{p2}
V&=&\frac12(1-\phi^{2})^2+2r^2\phi^2\chi^2+
\frac12r^2(\chi^2+\rho^2)^2+\nonumber\\
& &-r(1-\phi^2)(\chi^2+\rho^2)+2r^2(\phi-s)^2\rho^2
\een
It may support two, four, and six minima, depending on the values
of the two real parameters $r\neq0$ and $s\neq0$. In the case of six minima,
for $r>0$ and $-1<s<1$ we get
\ben
v_{1,2}&=&\left(\pm1, 0, 0 \right)
\\
v_{3,4}&=&\left(0,\pm\sqrt{\frac1r},0\right)
\\
v_{5,6}&=&\left(s,0,\mp\sqrt{\frac1r(1-s^2)}\;\right)
\een
The model engenders the $Z_2\times Z_2$ symmetry, corresponding to
reflections in the $\chi$ and $\rho$ axes. The presence of the $s$-dependent
term in the superpotential forbids reflection
symmetry in the $\phi$ axis. However, we notice that the operation
$\phi\to-\phi$ leads the system to a partner system, in which one
changes $s\to-s$.

Before studying the topological solutions of this system, it is
interesting to analyse the projections over diverse axis and planes.
The projection of the potential over the $\phi$ axis gives
\be
V(\phi,0,0)=\frac12(1-\phi^2)^2
\ee
which is known to produce kink-like domain walls that may host internal
structure.  In the $(\phi,\chi)$ plane we have
\ben
V(\phi,\chi,0)&=&\frac12(1-\phi^2)^2+2r^2\phi^2\chi^2\nonumber\\
& &-r(1-\phi^2)\chi^2+\frac12r^2\chi^4
\een
We see that inside the domain wall
\be
V(0,\chi,0)=\frac12r^2\left(\frac1{r}-\chi^2\right)^2
\ee
and that outside the wall
\be
V(\pm1,\chi,0)=2r^2\chi^2+\frac12r^2\chi^4
\ee
Thus, if the $\phi$ field gives rise to a domain wall, then inside this
wall the $\chi$ field generates another defect. 
Notice that the squared masses of the $\chi$ field  
inside and outside of the domain wall are given by $m^2_{\chi}(in)=4r$ and
$m^2_{\chi}(out)=4r^2$, showing that the parameter $r$ may
be used to control the presence of elementary $\chi$ mesons inside or
outside the host domain wall, according to $r$ being greater or smaller
than unit, respectively.

The projection of the potential in the $(\phi,\rho)$ plane gives
\ben
V(\phi,0,\rho)&=&\frac12(1-\phi^2)^2+2r^2(\phi-s)^2\rho^2\nonumber\\
& &-r(1-\phi^2)\rho^2+\frac12r^2\rho^4
\een
which takes the following forms, inside and outside the wall
\be
V(0,0,\rho)=\frac12+ r(2rs^2-1)\rho^2+\frac12r^2\rho^4
\ee
\be
V(\pm1,0,\rho)=2r^2(1\mp s)^2\rho^2+\frac12r^2\rho^4
\ee
Clearly there is no spontaneous symmetry breaking (SSB)
outside the host domain wall, but for $2rs^2<1$ there is SSB inside the
host domain wall and the system may generate domain walls with internal
structure. The mass of the $\rho$ meson inside the host wall,
in this case, reads $m^2_{\rho}(in)=4r(1-2rs^2)$. Outside the host wall
there is an asymmetry in the $+$ and $-$ sides which is controlled by the
vacua $v_{1,2}=(\pm1,0,0)$ and the mass of the $\rho$ meson is
$m^2_{\rho}(out,\pm)=4r^2(1\pm s)^2$. The $s$ parameter
induces the asymmetry for the $\rho$ meson, since its mass
depends on which side of the wall the meson is. Despite this asymmetry,
the ratio $m^2_{\rho}(in)/m^2_{\rho}(out,\pm)=(1/r)[(1-2rs^2)/(1\mp s)^2]$
shows that for $1/r<1-2|s|+3s^2$ the $\rho$ mesons prefer
to live inside the host wall. We notice that the two conditions $2rs^2<1$
and $1/r<1-2|s|+3s^2$ define a large region in parameter space, and
for $s$ very small we have $r>1+2|s|$. The special case $rs^2=1$ is also
worth mentioning since it defines an interesting region
in parameter space, in which there are explicit analytical solutions
connecting the minima $v_{3,4}$ to $v_{5,6}$ by linear orbits (see below).
In this case, as discussed above, the host wall cannot entrap topological
defects but the masses of the $\rho$ mesons can be explicitly computed to
give $m^2_{\rho}(in)=2/s^2$ inside the host wall and
$m^2_{\rho}(out,\pm)=4(1\mp s)^2/s^4$ outside. Thus,
$m^2_{\rho}(in)/m^2_{\rho}(out,\pm)=(1/2)[s/(1\mp s)]^2$,
showing that, even without explicit SSB but with $s$ sufficiently
small, the $\rho$ mesons prefer to live inside the host wall. The fact
that $\rho$ mesons prefer to live inside the wall is new, it does not
appear in the other case, in the plane $(\phi,\chi)$. More importantly,
although $s$ induces an asymmetry for the $\rho$ field, we may control
both $r$ and $s$ to entrap $\rho$ mesons inside the host wall, in a way
such that they do not perceive the asymmetry the model engenders.

The model defined by Eq.~(\ref{w2}) has several topological sectors.
All the sectors are BPS, except those connecting the minima
$v_{3}$ to $v_{4}$ and $v_{5}$ to $v_{6}$, which are of the non BPS type.
The energies of the BPS sectors are given as follows. In the $v_{1,2}$
minima sector we have $E_{BPS}^1=4/3$. In the sectors connecting
$v_{1,2}$ to $v_{3,4}$ we have $E_{BPS}^2=2/3$. For the connections
between $v_{1,2}$ and $v_{5,6}$
\be
\label{s1}
E_{BPS}^{\pm}(s)=\frac23\pm s\left(1-\frac{1}{3}s^2\right)
\ee
and for connections between $v_{3,4}$ and $v_{5,6}$
\be
\label{s2}
E_{BPS}(s)=|s|\left(1-\frac13 s^2\right)
\ee
We notice that these energies do not depend on $r$.
They are plotted below, to show their behavior with the parameter $s$.
Eq.~(\ref{s2}) shows that the limit $s\to0$
turns  to zero the energy in the sector connecting the minima
$(0,\pm\sqrt{1/r},0)$ and $(s,0,\pm\sqrt{(1-s^2)/r})$. This is
so because in the limit $s\to0$ these minima degenerate to a continuum
set of minima, which may be connected with no energy cost, according
to the Goldstone theorem. It is interesting to see that
$E^+_{BPS}(s>0)-E_{BPS}(s>0)$ and $E^-_{BPS}(s<0)-E_{BPS}(s<0)$ do not depend
on $s$, and that there are two specific values of $s$ [$s=\pm {\bar s}$,
with ${\bar s}^3-3{\bar s}+1=0$, $0<{\bar s}<1$]
for which $E_{BPS}(\pm{\bar s})=E^{\mp}_{BPS}(\pm{\bar s})$; for these
two values the energies of all the BPS sectors become equally
spaced, ordered as $E^1_{BPS}$, $E^+_{BPS}\; {\rm for}\; s>0\;
{\rm or}\;E^-_{BPS}\;{\rm for}\;s<0,\;E^2_{BPS},\;{\rm and}\;E^-_{BPS}\;
{\rm for}\;s>0\;{\rm or}\;E^+_{BPS}\;{\rm for}\;s<0$, with the values
$4/3,1,2/3,1/3$, respectively.

\begin{figure}[h]
\includegraphics[height=5.0cm,width=8.0cm]{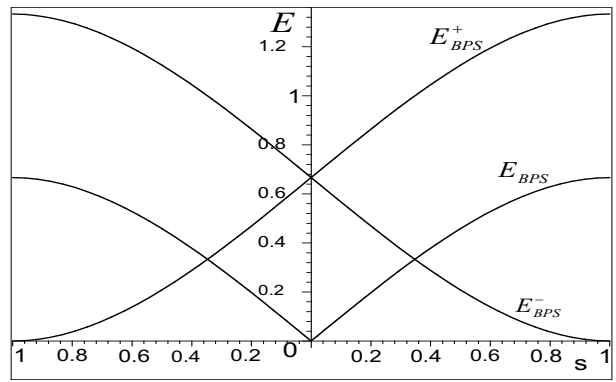}
\caption{Energies of the BPS states given by Eqs.~(\ref{s1})
and (\ref{s2}) as functions of the parameter s.}
\end{figure}

Let us now examine the presence of defects in the $s\neq0$ case.
The BPS states obey the first order equations
\ben
\label{foe}
\frac{d\phi}{dx}&=&1-\phi^2-r(\chi^2+\rho^2)
\\
\frac{d\chi}{dx}&=&-2r\phi\chi
\\
\frac{d\rho}{dx}&=&-2r(\phi-s)\rho
\een
In the sector connecting the minima $v_{1,2}$ there are topological
solutions that come from a straight line orbit with $\chi=\rho=0$,
which is described by $\phi(x)=\tanh(x)$. This is the one-field kink
solution for the system projected into the $\phi$ axis that generates
the domain wall hosting the other fields, as we referred to above.
For the system projected in the $(\phi,\chi)$ plane we get the
superpotential
\be\label{wxy}
W(\phi,\chi)=\phi-\frac13\phi^3-r\phi\chi^2
\ee
This model was first examined in \cite{b1}, and also in \cite{b2}.
In this case all the orbits in $(\phi,\chi)$ plane can be obtained
explicitly, as it was recently shown in Ref.~{\cite{es2}}.

For the system projected in the $(\phi,\rho)$ plane the
superpotential is
\be
W(\phi,\rho)=\phi-\frac13\phi^3-r(\phi-s)\rho^2
\ee
It gives rise to the first order equations
\ben
\frac{d\phi}{dx}&=&1-\phi^2-r\rho^2
\\
\frac{d\rho}{dx}&=&-2r(\phi-s)\rho
\een
To find solutions that connect the minima $v_{1,2}$ to $v_{5,6}$
we follow the approach developed in Ref.~{\cite{bfl}}. We found
straight line orbits connecting these minima with
\be
\label{sl1}
\rho=\pm\sqrt{\frac1{r}\frac{1+s}{1-s}}\,(1-\phi)
\ee
and 
\be
\label{sl2}
\rho=\pm\sqrt{\frac1{r}\frac{1-s}{1+s}}\,(1+\phi)
\ee
For the first trajectories, Eq.(\ref{sl1}), there are BPS states
for $1/r=1-s$. They are
\ben
\phi(x)&=&\frac12(1+s)+\frac12(1-s)\tanh(x)
\\
\rho(x)&=&\pm\frac12(1-s)\sqrt{1+s}[1-\tanh(x)]
\een
They obey $\phi(x\to-\infty)\to s$ and $\rho(x\to-\infty)\to(1-s)\sqrt{1+s}$,
and $\phi(x\to\infty)\to1$ and $\rho(x\to-\infty)\to0$. Thus, they connect
$(s,0,\pm\sqrt{(1-s^2)(1-s)})$ to $(1,0,0)$. For the second trajectories,
Eq.(\ref{sl2}), there are BPS states for $1/r=1+s$. They are
\ben
\phi(x)&=&-\frac12(1-s)+\frac12(1+s)\tanh(x)
\\
\rho(x)&=&\pm\frac12(1+s)\sqrt{1-s}[1+\tanh(x)]
\een
They connect $(-1,0,0)$ to $(s,0,\pm\sqrt{(1-s^2)(1+s)})$.

There are other orbits connecting the minima $v_{1,2}$ and $v_{5,6}$.
We found the following parabolic trajectories
\be
\rho^2=\frac1{2r}(1-s^2)(1\pm\phi)
\ee
giving rise to BPS states connecting the minima for $r=1/4$.
They have the form
\ben
\phi(x)&=&\frac12(1+s)+\frac12(1-s)\tanh[x/L^{-}(s)]
\\
\rho(x)&=&\pm[2(1-s^2)]^{1/2}
\{1-\tanh[x/L^{-}(s)]\}^{1/2}
\een
and
\ben
\phi(x)&=&-\frac12(1-s)+\frac12(1+s)\tanh[x/L^{+}(s)]
\\
\rho(x)&=&\pm[2(1-s^2)]^{1/2}
\{1+\tanh[x/L^{+}(s)]\}^{1/2}
\een
where $L^{\pm}(s)=2/(1\pm s)$ give the width of the topological defects.

We also found three-field BPS states connecting the minima $v_{3,4}$ to
$v_{5,6}$. This is even harder, because no field can vanish along any
smooth orbit connecting $v_{3,4}$ to $v_{5,6}$. We first notice
that there are four different ways to connect $(0,\pm\sqrt{1/r},0)$
and $(s,0,\pm\sqrt{(1-s^2)/r})$. We found straight line orbits described by
\be
\phi\pm s\sqrt{r}\chi-s=0
\ee
and
\be
\sqrt{\frac1{r}(1-s^2)}\,\phi\mp s\rho=0
\ee
which imply that
\be
\rho+\sqrt{1-s^2}\chi\mp\sqrt{\frac1{r}(1-s^2)}=0
\ee
These orbits also project as straight line segments
in each one of the three planes $(\phi,\chi)$, $(\phi,\rho)$ and
$(\chi,\rho)$. We use them to rewrite Eq.~(\ref{foe}) in the form
\be
\frac{d\phi}{dx}=\frac2{s}\left(\phi-\frac1{s}\phi^2\right)
\ee
which is solved by
\be
\phi(x)=\frac{s}{2}[1+\tanh(x/s)]
\ee
The other solutions are
\ben
\chi(x)&=&\pm\frac12\sqrt{\frac1{r}}[1-\tanh(x/s)]
\\
\rho(x)&=&\pm\frac12\sqrt{\frac1{r}(1-s^2)}[1+\tanh(x/s)]
\een
They require that $rs^2=1$.

As we have shown, we have been able to find explicit BPS solutions
in all the BPS sectors of a model described by three real scalar field
with discrete symmetry. The presence of the parameter $s$ restricts
the symmetry of the model to $Z_2\times Z_2$, since $s\neq0$ forbids
reflection symmetry along the $\phi$ axis. For this reason, now we
study the behavior of the system in the limit $s\to0$. In this case the
superpotential (\ref{w2}) becomes
\be
\label{w1}
W=\phi-\frac13\phi^3-r\phi(\chi^2+\rho^2)
\ee
The corresponding potential supports the two minima $(\pm1,0,0)$, and also
$(0,\sqrt{1/r}\cos\theta,\sqrt{1/r}\sin\theta)$, which defines a circle
in the $(\chi,\rho)$ plane for $\theta\in[0,2\pi)$.
The first order equations are
\ben\label{bc1}
\frac{d\phi}{dx}&=&1-\phi^2-r(\chi^2+\rho^2)
\\
\frac{d\chi}{dx}&=&-2r\phi\chi\label{bc2}
\\
\frac{d\rho}{dx}&=&-2r\phi\rho\label{bc3}
\een
There are solutions connecting the minima $(\pm1,0,0)$. One of them is
given by $\chi=\rho=0$ and $\phi(x)=\tanh(x)$. Its energy is $4/3$,
and it may give rise to a domain wall in $(3,1)$ space-time dimensions.
There are other solutions connecting the minima $(\pm1,0,0)$, with the
same energy of the above solution. They are given by
\ben\label{s01}
\phi(x)&=&\tanh(2rx)
\\
\label{s02}
\chi(x)&=&\pm\sqrt{\frac1{r}-2\;}\frac{\cos\theta}{\cosh(2rx)} 
\\
\label{s03}
\rho(x)&=&\pm\sqrt{\frac1{r}-2\;}\frac{\sin\theta}{\cosh(2rx)}
\een
They form an elliptical surface, and in Fig.~[2] we depict these solutions
in the case $r=1/6$ and $\theta=\pi/6$.

\begin{figure}[h]
\includegraphics[height=5.0cm,width=8.0cm]{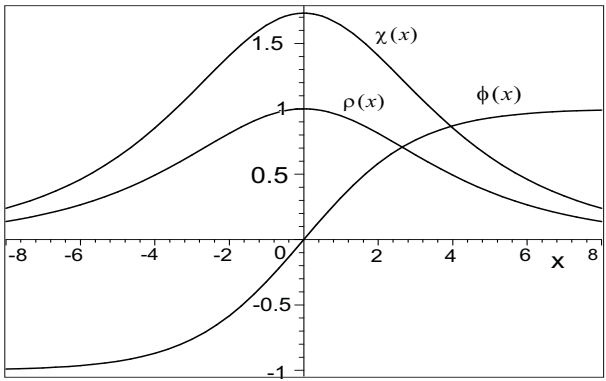}
\caption{The BPS solution described by Eqs.~(\ref{s01}),
(\ref{s02}), and (\ref{s03}), for $r=1/6$ and $\theta=\pi/6$.}
\end{figure}

We notice that this three-field solution may be used to describe three-wave
parametric solitons in quadratic nonlinear media \cite{pha}. Moreover, the
fact that $\chi$ and $\rho$ are non-vanishing for $x\to0$ indicates that the
wall generated by the $\phi$ field may entrap both $\chi$ and $\rho$ in its
interior -- see Refs.~{\cite{is1,is2,00a}} for further details.

There are other solutions, connecting the minima $(\pm1,0,0)$ to
$(0,\sqrt{1/r}\cos\theta,\sqrt{1/r}\sin\theta)$ through parabolic surfaces.
Some of them are, for $r=1/4$, and for $\chi^2+\rho^2=(1/4)(1-\phi)$
\ben
\phi(x)&=&\frac12(1+\tanh(x/2))
\\
\chi(x)&=&\pm\cos\theta\sqrt{2(1-\tanh(x/2))}
\\
\rho(x)&=&\pm\sin\theta\sqrt{2(1-\tanh(x/2))}
\een
and also, for $\chi^2+\rho^2=(1/4)(1+\phi)$
\ben
\phi(x)&=&-\frac12(1-\tanh(x/2))
\\
\chi(x)&=&\pm\cos\theta\sqrt{2(1+\tanh(x/2))}
\\
\rho(x)&=&\pm\sin\theta\sqrt{2(1+\tanh(x/2))}
\een
These solutions have the same energy, given by the value $2/3$. This is
one-half the value of the energy in the former sector, defined by the
minima $(\pm1,0,0)$.

We notice that the potential generated by $W(\phi,\chi,\rho)$ as in
Eq.~(\ref{w1}) is such that
\be
V(0,\chi,\rho)=\frac12r^2\left(\frac1r-\chi^2-\rho^2\right)^2
\ee
It shows that the domain wall generated by the $\phi$ field may entrap
the other two fields in its interior, giving rise to a model where one can
have SSB of the continuum, global $U(1)$ symmetry. In Eqs.~(\ref{bc1}),
(\ref{bc2}), and (\ref{bc3}), if we change $\chi\to\sigma \cos\theta$
and $\rho\to\sigma\cos\theta$, for $\theta$ constant we get to the
equations $d\phi/dx=1-\phi^2-r\sigma^2$ and $d\sigma/dx=-2r\phi\sigma$,
which are similar to the first order equations that appear in
the model represented by Eq.~(\ref{wxy}), thus we  can follow
Ref.~{\cite{es2}} to find all the BPS states of the model.

The continuum, global $U(1)$ symmetry that appear for $s=0$ can
be further explored. For instance, we notice that in $(3,1)$ space-time
dimensions the host wall $\phi(x)$ depends only on $x$, and may entrap
the other two fields in two spatial dimensions. Now, if we let the continuum
symmetry that appear in the $(\chi,\rho)$ plane to
become local, and if we start with all the needed physical ingredients,
the model gives rise to a host domain wall that entraps
a planar system, driving us for instance to the scenario where the so-called
Callan-Harvey effect \cite{ch} may spring very naturally. We recall that
the Callan-Harvey effect concerns anomaly cancelation in $(2,1)$ space-time
dimensions due to the presence of a domain wall mass to a fermionic field,
and we quote Ref.~{\cite{cha}} for an Abelian version of this effect.
This issue is presently under investigation, and we hope to report
on it in the near future.

Another line of investigation concerns the high temperature effects,
which may drive the system to phase transitions. This line may follow
the two last work in Ref.~{\cite{is2}}, which deal with the thermal effects
in the case of two real scalar fields. Furthermore, since the high temperature
 effects are obtained {\it a la} Matsubara, which is a
compactification process, we can also consider more general
possibilities \cite{mms}, in which one investigates the
effective potential in the case two or more dimensions
of the original space-time are compactified. Similar ideas have been
introduced in \cite{rwo} for bosons and fermions, examining the
temperature inversion symmetry. This direction brings other
issues since together with the usual thermal effects, we are also
investigating effects of compactification of one or more spatial
dimensions.

The presence of topological sectors depends on the manifold of vacuum states
of the system, and this indicates that our investigations are model-dependent.
Thus we can build other three-field models, exploring the possibility of
finding spatial junctions of defects, which may allow that we fill space
with solid structures, generalizing the idea of tiling the plane with
regular poligons. In the plane there are an infinity of regular polygons,
although only three of them are capable of tiling the plane.
In space, however, there are only five regular polyhedra -- the five Platonic
solids -- and the issue is then which are capable of filling space, and how
we can generate them within the context of three-field models. These and
other related issues are presently under consideration, and we hope to
report on them in the near future.

We would like to thank A.R. Gomes, J.R.S. Nascimento and R.F. Ribeiro
for discussions, and CAPES, CNPq, PROCAD and PRONEX for partial support.

\end{document}